\documentstyle[pre,aps,twocolumn,amsfonts,amssymb,eqsecnum]{revtex}
\begin{document}
\def\u{\bbox}
\def\Bbb{\relax}
\def\d{\displaystyle}
\def\mathcal#1{{\cal #1}}
\def\goldenmean{\gamma}
\def\phi{\varphi}
\def\epsilon{\varepsilon}
\def\goldenmean{\gamma}
\newtheorem{lemme}{Lemma}
\newtheorem{theorem}{Theorem}

\draft

\title{A Version of Thirring's Approach to the KAM Theorem 
for Quadratic Hamiltonians with Degenerate Twist}

\author{C. Chandre and H. R. Jauslin}

\address{Laboratoire de Physique, CNRS, Universit\'e de Bourgogne,
BP 400, F-21011 Dijon, France}
\maketitle

\begin{abstract}
We give a proof of the KAM theorem on the existence of 
invariant tori for weakly perturbed Hamiltonian systems, 
based on Thirring's approach for Hamiltonians that are
quadratic in the action variables. The main point of this approach is 
that the iteration of canonical transformations on which the proof is 
based stays within the space of quadratic Hamiltonians. 
 We show that Thirring's proof for 
nondegenerate Hamiltonians can be adapted to Hamiltonians with degenerate
twist.
This case, in fact, drastically simplifies Thirring's proof.
\end{abstract}

\section{Introduction}

Regular Hamiltonian dynamics is characterized by the existence of as many
independent conserved quantities as degrees of freedom $d$. As a
consequence, each trajectory is confined to evolve on an invariant torus
of dimension $d$. The KAM technique was developed to prove the 
stability under small perturbations of a large fraction of these invariant tori.
The KAM theorem\cite{kolmogorov,arnold,moser,gallavotti3}
states that, if the frequency 
satisfies a diophantine condition, and the size of the perturbation 
$\varepsilon$ is sufficiently small, 
then a torus of that frequency will be stable.
The proof is based on an iterative algorithm to construct the invariant 
tori. Each step of the KAM iteration consists of a coordinate
transformation that reduces the size of the perturbation from order
$\varepsilon$ to $\varepsilon^2$.\\
In this paper, we derive a KAM theorem
for a family of Hamiltonians that has been used in Refs.\
\cite{govin,chandre,cgjk} within the setup of a
renormalization-group approach to the breakup of invariant tori.\\
We consider the following class of Hamiltonians
with $d$ degrees of freedom, that are
quadratic in the action variables $\u{A}=(A_1,A_2,\ldots,A_d)\in{\Bbb R}^d$ 
and described 
by three scalar functions of the angles ${\u \varphi}=(\varphi_1,
\varphi_2,\ldots,\varphi_d)\in{\Bbb T}^d$ (the $d$-dimensional torus 
parametrized, e.\ g. by $[0,2\pi]^d$):
\begin{eqnarray}
H({\u A},{\u \varphi})=&&
\frac{1}{2}m({\u \varphi})( {\u \Omega}\cdot{\u A})^{2} \nonumber \\
&& +\lbrack {\u \omega}_{0}
+ g({\u \varphi}){\u \Omega} \rbrack \cdot{\u A}
+ f({\u \varphi}) \ , \label{hamiltonian}
\end{eqnarray}
where ${\u \omega}_0\in {\Bbb R}^d$ 
is the frequency vector of the considered torus, and ${\u \Omega}$ 
is some other constant vector. Without loss of generality, we assume 
that ${\u \Omega}=(\Omega_1,\ldots,\Omega_d)$ is of norm one, i.\ e.\ 
$\vert {\u \Omega}\vert =\sum_{i=1}^d\vert \Omega_i\vert =1$.
The functions $m$, $g$, and $f$ are real analytic
on ${\Bbb T}^d$, i.\ e.\ they are holomorphic functions on some
complex neighborhood of ${\Bbb T}^d$.\\
We consider ${\u \omega}_0$ verifying a diophantine condition
\begin{equation}
  \label{diophan}
  \vert{\u \omega}_0\cdot{\u \nu}\vert^{-1} \leq \sigma 
  \vert{\u \nu}\vert^{\tau},
  \; \forall {\u \nu}=(\nu_1,\ldots,\nu_d)\in {\Bbb Z}^d \setminus\{0\},
\end{equation}
for some $\tau > d-1$ and $\sigma>0$.\\
The aim is to prove the existence of a torus with frequency vector 
${\u \omega}_0$ for Hamiltonian systems described by Eq.\ (\ref{hamiltonian}).
The method is to find a canonical transformation such that the equations
of motion expressed in the new coordinates show trivially
the existence of this torus. For Hamiltonians 
of type (\ref{hamiltonian}), if one
takes $g$ and $f$ equal to zero, then the resulting equations of motion are
\begin{eqnarray}
 && \frac{d{\u A}}{dt}=-\d\frac{1}{2}\frac{\partial m}{\partial {\u \phi}}
  ({\u \Omega} \cdot {\u A})^2, \\
&& \frac{d{\u \phi}}{dt}=
  m({\u \phi})({\u \Omega} \cdot {\u A}) {\u \Omega}+{\u \omega}_0.
\end{eqnarray}
Thus, there exists a torus with frequency vector $\u{\omega}_0$ located at
$\u{A}=0$ (even if the resulting Hamiltonian is not globally integrable,
i.\ e. for ${\u A}\not= 0$).
This canonical transformation cannot be defined directly, because the
formal expressions that appear in the classical perturbation theory do
not converge due to the presence of small denominators. The construction
is done via an iterative algorithm. We iterate a canonical change of coordinates
that maps a Hamiltonian of the form (\ref{hamiltonian}) 
with a perturbation $(g,f)$ of order 
$\mathcal{O}(\varepsilon)$, into a Hamiltonian with $(g',f')$ of order
$\mathcal{O}(\varepsilon^2)$. When the iteration converges, the perturbation
$(g,f)$ is completely eliminated.\\
The fact that $m$ does not need to be eliminated to prove the existence of
the torus with frequency vector ${\u \omega}_0$, allows us to stay
with Hamiltonians that are quadratic in the actions at each step of 
the iteration. This approach has been developed by 
Thirring~\cite{thirring,jauslin}.
In fact, Thirring considered a non-degenerate family
of Hamiltonians quadratic in the actions, of the form
\begin{eqnarray}
H({\u A},{\u \varphi})=&&
\frac{1}{2}\u{A}\cdot M({\u \varphi})\u{A} \nonumber \\
&& +\lbrack {\u \omega}_{0}
+ \u{g}({\u \varphi}) \rbrack \cdot{\u A}
+ f({\u \varphi}) \ , \label{hamiltonian2}
\end{eqnarray}
where $M$ is a $d\times d$ matrix such that $\mbox{det } M\not=0$,
and $\u{g}$ a vector. This implies that Thirring's Hamiltonian 
(\ref{hamiltonian2}) satisfies the ``twist condition''
\begin{equation}
\label{eqn:twist}
\mbox{det }\left| \frac{\partial^2 H}{\partial \u{A} \partial \u{A}}
\right| = \mbox{ det } M \not= 0.
\end{equation}
The Hamiltonian (\ref{hamiltonian}) 
 does not satisfy the twist condition for any ${\u \varphi}$. The extension of
the KAM theorem for degenerate systems was done by Arnold~\cite{arnold2}
(see also Refs.\ \cite{chengsun,russmann,gallavotti1,gallavotti2,paul}).
In the case we consider here, the rank of the matrix in (\ref{eqn:twist})
is one. Thus, there is a twist in a particular direction, which is
the only relevant one for the considered perturbation.\\
Condition (\ref{eqn:twist}) is also called (standard) nondegeneracy 
condition. KAM theorems were also proven for Hamiltonians which 
satisfy the isoenergetic 
nondegeneracy condition~\cite{delshams,broer},
for which the following determinant of order $d+1$ does not vanish 
\begin{equation}
\label{cond:iso}
\mbox{det }\begin{array}{|cc|} \d \frac{\partial^2 H}{\partial \u{A} \partial
\u{A}}
& \d \frac{\partial H}{\partial \u{A}} \\
& \\
\left(\d \frac{\partial H}{\partial \u{A}}\right)^T & 0 \end{array} \not= 0.
\end{equation}
For $d=2$, the determinant
(\ref{cond:iso}) is equal to $-m({\u \varphi}) [\mbox{det }({\u \Omega},
{\u \omega}_0)]^2$. Thus, the isoenergetic nondegeneracy condition is not
satisfied if $m({\u \varphi})$ has zeroes.\\
For $d>2$, the model (\ref{hamiltonian}) is isoenergetically
degenerate. The present result is thus not
a direct consequence of the ones in \cite{delshams,broer}.
We notice that in the case where ${\u \Omega}$
is parallel to ${\u \omega}_0$, the Hamiltonian (\ref{hamiltonian}) is
integrable (because of the existence of $d$ integrals of motion
in involution : $H$ and ${\u \omega}_0^{\perp}\cdot {\u \varphi}$,
where ${\u \omega}_0^{\perp}$ denotes a vector perpendicular to
${\u \omega}_0$, and there are $d-1$ such independant vectors).\\
In this article, we present a self-contained proof of the KAM theorem 
for Hamiltonian 
(\ref{hamiltonian}) based on Thirring's proof for Hamiltonian 
(\ref{hamiltonian2}). The advantages are twofold: On one hand, we show
that Thirring's proof can be adapted to degenerate twist Hamiltonians, and on 
the
other hand, the resulting proof becomes even simpler than Thirring's. 
The fact that the iteration stays within the space of Hamiltonians 
quadratic in the actions is very useful, e.\ g., for numerical 
implementations to study the breakup of invariant tori 
\cite{govin,chandre,cgjk}.\\
Before entering into details of the theorem, we give basic definitions and
notations: We denote $\mathcal{D}_\rho$ a complex neighborhood of 
${\Bbb T}^d$ defined by
\begin{equation}
  \mathcal{D}_\rho=\{ {\u \varphi}\in {\Bbb C}^d \vert \quad
  \Vert \mbox{Im } {\u \varphi} \Vert \leq \rho \},
\end{equation}
where $\Vert z \Vert=\max_i(\vert z_i\vert)$, 
for any $z\in {\Bbb C}^d$. We will consider, in the following calculations,
scalar functions defined in $\mathcal{D}_\rho$. More precisely, we define
$\mathcal{A}_\rho$ as the set of complex
functions $f({\u \varphi})$
defined on $\mathcal{D}_\rho$, analytic in the interior of $\mathcal{D}_\rho$,
of period $2\pi$ in the variables $\varphi_i$, and which have real values 
when ${\u \varphi}\in {\Bbb R}^d$. We define a norm on $\mathcal{A}_\rho$:
$\Vert f \Vert_\rho=\d {\mathrm sup}_{\varphi \in \mathcal{D}_\rho} 
\vert f({\u \varphi}) \vert$.
We define $\langle f \rangle$, the mean value of $f$ by
\begin{equation} 
\langle f \rangle=\int_{ {\Bbb T}^d}\frac{d^d\u{\phi}}{(2\pi)^d} f(\u{\phi}).
\end{equation}
In the following sections, 
we will use the notation $\u{\partial}f=\d \frac{\partial f}{\partial
\u{\varphi}}$ for any function of the angles.\\
In Sec.\ \ref{sec:expr}, we define the KAM iteration. In Sec.\ \ref{sec:esti},
we give estimates on the transformed functions. Finally, in 
Sec.\ \ref{sec:conv},
we iterate the transformation, and prove the following 
KAM theorem for the  family of 
Hamiltonians (\ref{hamiltonian}):
\begin{theorem} 
For $H({\u A},{\u \varphi})=\d
       \frac{1}{2}m({\u \varphi})( {\u \Omega}\cdot{\u A})^{2}
       +\lbrack {\u \omega}_{0}
       + g({\u \varphi}){\u \Omega} \rbrack \cdot{\u A}
       + f({\u \varphi}) $,
       suppose that \\
\indent $(i)$ $\vert{\u \omega}_0\cdot{\u \nu}\vert^{-1} \leq 
\sigma \vert{\u \nu}\vert^{\tau},
              \; \forall {\u \nu}\in {\Bbb Z}^d \setminus\{0\}$, 
	      for some $\sigma>0$ and $\tau>d-1$;\\
\indent $(ii)$ $m$, $g$, and $f$ are analytic in $\mathcal{D}_\rho$, 
and $\vert {\u \Omega} \vert = 1$;\\
\indent $(iii)$ $\max(\Vert g\Vert_\rho,\Vert f\Vert_\rho)\leq
\beta_0,$
where \begin{eqnarray*}
        && \beta_0=2^{-10}\Gamma^{-6} c^{-3} (h/3)^{3(\tau+d+1)},\\
        && c=2^{3d} \sigma
                     \left[2e^{-1}(\tau +1)\right]^{\tau+1},\\
	&& \Gamma=\max(1,\Vert m\Vert_\rho,\vert \langle m\rangle \vert^{-1}),\\
	&& h<\inf(2\rho/9,c^{1/(\tau+d+1)});
      \end{eqnarray*}
Then, there exists a canonical transformation, analytic in $\mathcal{D}_{\rho-
9h/2}$, such that the Hamiltonian expressed in the new coordinates is
$$ H^{(\infty)}=\frac{1}{2}m^{(\infty)}({\u \varphi}^{(\infty)})
         ({\u \Omega}\cdot{\u A}^{(\infty)})^{2}
       + {\u \omega}_{0}\cdot{\u A}^{(\infty)},
$$
where $m^{(\infty)}$ is analytic in $\mathcal{D}_{\rho-9h/2}$.
 As a consequence, the system has an analytic invariant torus of 
frequency ${\u \omega}_{0}$. 
\end{theorem}
Remark: if the constant part $\langle m\rangle$ of the quadratic term is
zero, the maximal amplitude of the perturbation $\beta_0$ given by the 
theorem becomes zero. Thus, in order to have a nontrivial result, we
require that $\langle m\rangle$ is nonzero.

\section{Expressions of the generating function and of the new Hamiltonian}
\label{sec:expr}

We perform a canonical transformation
$\mathcal{U}_F:(\u{\phi},\u{A})\mapsto (\u{\phi}',\u{A}')$ 
defined by a generating function~\cite{goldstein,gallavottiL} linear in the
action variables, and characterized by two
scalar functions $Y$, $Z$, of the angles, and a constant $a$, of the form
\begin{equation}
\label{eqn:Scan}
F(\u{A}',\u{\phi})=\left(\u{A}'+a\u{\Omega}\right)\cdot\u{\phi}
+ Y(\u{\phi})\u{\Omega}\cdot\u{A}'+ Z(\u{\phi}),
\end{equation}
leading to
\begin{eqnarray}
&&  \label{eqn:tcan1}
    \u{A}=\d \frac{\partial F}{\partial \u{\phi}}
	 =\u{A}'+\left( \u{\Omega}\cdot\u{A}'\right) \u{\partial}Y
                  + a \u{\Omega} 
                  + \u{\partial} Z,\\
&& \label{eqn:tcan2}
    \u{\phi}'=\d \frac{\partial F}{\partial \u{A}'}
             =\u{\phi}+Y(\u{\phi})\u{\Omega}.
\end{eqnarray}
Inserting Eq.\ (\ref{eqn:tcan1}) into the Hamiltonian (\ref{hamiltonian}),
we obtain the 
expression of the Hamiltonian in the mixed representation
of new action variables and old angle variables
\begin{eqnarray}
\tilde{H}({\u A}',{\u \varphi})=&&
\frac{1}{2}\tilde{m}({\u \varphi})( {\u \Omega}\cdot{\u A}')^{2} \nonumber \\
&& +\lbrack {\u \omega}_{0}
+ \tilde{g}({\u \varphi}){\u \Omega} \rbrack \cdot{\u A}'
+ \tilde{f}({\u \varphi}) \ , \label{hamaction}
\end{eqnarray}
with
\begin{eqnarray}
 && \tilde{m}=(1+\u{\Omega}\cdot\u{\partial}Y)^2 m,\\
 && \tilde{g}=g+\u{\omega}_0\cdot\u{\partial}Y+m b+
            \u{\Omega}\cdot\u{\partial}Y\left(g+m b\right),\\
 && \tilde{f}=f+\u{\omega}_0\cdot\u{\partial}Z+\frac{1}{2}m b^2 +g b,
 \label{eqn:tcanf}
\end{eqnarray}
where $b(\u{\phi})=a\Omega^2+\u{\Omega}\cdot\u{\partial}Z$. 
As the new angles do not depend on the actions, 
the Hamiltonian (\ref{hamiltonian})
expressed in the new variables is also quadratic in the actions, and of the
same form as (\ref{hamiltonian}).
We notice that 
this transformation does not change $\u{\Omega}$.\\
We determine the generating function (\ref{eqn:Scan})
such that the new functions $\tilde{g}$, $\tilde{f}$ in
$\mathcal{H}\circ\mathcal{U}_F$ 
vanish to the first order in $\epsilon$.
This leads to the conditions
\begin{eqnarray}
   && \label{eqn:cond1}
   \u{\omega}_0\cdot\u{\partial}Z+f=const,\\
   && \label{eqn:cond2}
   \u{\omega}_0\cdot\u{\partial}Y+g
   +m\left(a\Omega^2+\u{\Omega}\cdot\u{\partial} Z\right)=0.
\end{eqnarray}
The constant $a$ allows us to have $\langle \tilde{g}\rangle=\mathcal{O}(
\varepsilon^2)$, in order to keep the frequency ${\u \omega}_0$ at the
chosen value.
We recall that the functions $g$ and $f$ are of order 
$\mathcal{O}(\epsilon)$ and $m$ is of order one; 
as a consequence $Y$, $Z$ and $a$ are of order $\mathcal{O}(\epsilon)$.
Equations (\ref{eqn:cond1}) and (\ref{eqn:cond2}) are solved by representing 
them 
in Fourier space. They define the generating function $F$ as
\begin{equation}
 \label{eqn:Z}
  Z(\u{\phi})=\sum_{\nu\not= 0} \frac{i}{\u{\omega}_0\cdot\u{\nu}}
                 f_{\nu} e^{i\nu\cdot\phi},
\end{equation}
\begin{equation}
\label{eqn:a}
  a=-\frac{\langle g\rangle+\langle m\u{\Omega}\cdot\u{\partial}Z\rangle}
          {\Omega^2\langle m \rangle},
\end{equation}
\begin{eqnarray}
  Y(\u{\phi})=\sum_{\nu\not= 0} \frac{i}{\u{\omega}_0\cdot\u{\nu}}
                 &&\left(g_{\nu}+(m\u{\Omega}\cdot\u{\partial}Z)_{\nu} \right.
		 \nonumber \\
		&&\quad \left. +m_{\nu} a \Omega^2\right)
                  e^{i\nu\cdot\phi}, \label{eqn:Y}
\end{eqnarray}
where the scalar functions $m$, $g$, $f$ are represented by their Fourier
series, e.\ g.
\begin{equation}
  f({\u \varphi})=\sum_{\nu \in {\Bbb Z}^d} f_\nu e^{i\nu\cdot\varphi}.
\end{equation}
Thus the scalar functions in $\tilde{H}$ become
\begin{eqnarray}
 && \label{eqn:m}
 \tilde{m}=(1+\u{\Omega}\cdot\u{\partial}Y)^2 m,\\
 && \label{eqn:g}
 \tilde{g}=-\left(\u{\omega}_0\cdot\u{\partial}Y\right)
            \left(\u{\Omega}\cdot\u{\partial}Y\right),\\
 && \label{eqn:f}
 \tilde{f}=\frac{1}{2}(g-\u{\omega}_0\cdot\u{\partial}Y)
              \left(a\Omega^2+\u{\Omega}\cdot\u{\partial}Z\right).
\end{eqnarray}
The expression of $\tilde{H}$ in the new angles requires the inversion of 
Eq.\ (\ref{eqn:tcan2}). We denote $H'$ the Hamiltonian expressed in the 
new variables $H'({\u A}',{\u \varphi}')=\tilde{H}({\u A}',{\u \varphi})$:
\begin{eqnarray}
H'({\u A}',{\u \varphi}')=&&
\frac{1}{2}m'({\u \varphi}')( {\u \Omega}\cdot{\u A}')^{2} \nonumber \\
&&+\lbrack {\u \omega}_{0}
+ g'({\u \varphi}'){\u \Omega} \rbrack \cdot{\u A}'
+ f'({\u \varphi}') \ .
\end{eqnarray}
In the following section, we give estimates on $Z$, $Y$ and $a$, 
in order to derive estimates on $m'$, $g'$ and $f'$.

\section{Estimates on the generating function and on the new Hamiltonian}
\label{sec:esti}

  \subsection{Estimate on the generating function}
  
\begin{lemme} 
Let $f$ be element of $\mathcal{A}_\rho$, and $\u{\omega}_0$ satisfy
Eq.\ (\ref{diophan}), then $Z(\u{\varphi})$ given by Eq.\ (\ref{eqn:Z})
is analytic on $\mathcal{D}_{\rho-h}$
(for any choice of $h$ with $0<h<\rho<2$), 
and satisfies the following estimates
\begin{eqnarray}
&& \Vert Z \Vert_{\rho - h} \leq \bar{c} h^{-\tau-d} \Vert f \Vert_\rho,
   \label{eqn:estZ} \\
&& \Vert {\u \omega}_0\cdot {\u \partial} Z \Vert_{\rho - h} \leq
   \Vert f \Vert_\rho,\label{eqn:estDZ1} \\
&& \Vert {\u \Omega}\cdot {\u \partial} Z \Vert_{\rho - h} \leq
   c h^{-\tau-d-1}\Vert f \Vert_\rho,\label{eqn:estDZ2}
\end{eqnarray}
where $\bar{c}=2^{3d}\sigma\left(\d 2e^{-1}\tau\right)^\tau$ and 
$c=2^{3d} \sigma
        \left[2e^{-1}(\tau +1)\right]^{\tau+1}$.
\end{lemme}
\emph{Proof:}\\
This lemma is proved, for instance, in Refs.\ \cite{arnold2,benettin}.\\
First, we estimate the Fourier coefficients of $f$ expressed as integrals 
of $f$ over the torus,
\begin{equation}
f_\nu=\int_{{\Bbb T}^d} \frac{d^d{\u \varphi}}{(2\pi)^d} f({\u \varphi})
e^{-i\nu\cdot\varphi}.
\end{equation}
By a shift of the angles ${\u \varphi}\mapsto{\u \varphi}+i{\u \eta}$,
where $\eta_i=-\rho \nu_i/\vert \nu_i\vert$,
\begin{equation}
f_\nu=\int_{{\Bbb T}^d} \frac{d^d{\u \varphi}}{(2\pi)^d} 
f({\u \varphi}+i{\u \eta})
e^{-i\nu\cdot\varphi} e^{-\rho\vert \nu\vert}.
\end{equation}
Then $\vert f_\nu \vert \leq \Vert f\Vert_\rho e^{-\rho\vert \nu\vert}$, for
all ${\u \nu}\in {\Bbb Z}^d$.\\
To estimate $Z$ given by Eq.\ (\ref{eqn:Z}), we use the diophantine property
(\ref{diophan}),
\begin{equation}
\vert Z({\u \varphi})\vert \leq \sigma\sum_{\nu\not= 0} 
\vert {\u \nu}\vert^\tau 
\vert f_\nu\vert e^{-\nu\cdot {\mathrm Im}\varphi}.
\end{equation}
Then, for all ${\u \varphi}\in \mathcal{D}_{\rho-h}$, we have
\begin{equation}
\vert Z({\u \varphi})\vert \leq \sigma\Vert f\Vert_\rho \sum_{\nu\not= 0} 
\vert {\u \nu}\vert^\tau e^{-h\vert \nu\vert}.
\end{equation}
To estimate the sum, we use the following property which is easy to check
(see Ref.\ \cite{arnold}):
\begin{equation}
\vert {\u \nu}\vert^\tau \leq \left(\frac{2\tau}{eh}\right)^\tau 
e^{\vert \nu \vert h/2} \, , \quad \forall \tau,h >0.
\end{equation}
From the fact that $1/(1-e^{-x})<2/x$, for all $x\in ]0,1[$, we have 
\begin{equation}
\sum_{\nu\not= 0} e^{-\vert \nu\vert h/2} < \left(\frac{8}{h}\right)^d,
\end{equation}
which gives the estimate (\ref{eqn:estZ}).
The estimate (\ref{eqn:estDZ2}) is obtained by the same calculations,
 and the estimate
(\ref{eqn:estDZ1}) is straightforward from Eq. (\ref{eqn:cond1}). 
 $\Box$ \\

\begin{lemme}
Let $m$, $g$, and $f$ be elements of $\mathcal{A}_\rho$, 
and $\u{\omega}_0$ satisfy
Eq.\ (\ref{diophan}), then $Y(\u{\varphi})$ given by 
Eq.\ (\ref{eqn:Y})
is analytic on $\mathcal{D}_{\rho-2h}$. The constant $a$ and the function $Y$
satisfy the following estimates
\begin{eqnarray}
\Vert Y \Vert_{\rho - 2h} \leq \bar{c} h^{-\tau-d} && \left(1+\vert \langle m 
\rangle \vert^{-1} \Vert m \Vert_\rho \right) \nonumber \\
&&\times \left( \Vert g\Vert_\rho +
c h^{-\tau-d-1} \Vert f\Vert_\rho \, \Vert m\Vert_\rho \right),\label{eqn:estY}
\end{eqnarray}
\begin{eqnarray}
\Vert {\u \omega}_0 \cdot {\u \partial} Y \Vert_{\rho - 2h} 
\leq && \left(1+\vert \langle m 
\rangle \vert^{-1} \Vert m \Vert_\rho \right) \nonumber \\
&& \times \left( \Vert g\Vert_\rho +
c h^{-\tau-d-1} \Vert f\Vert_\rho \, \Vert m\Vert_\rho \right),
\label{eqn:estDY1}
\end{eqnarray}
\begin{eqnarray}
\Vert {\u \Omega} \cdot {\u \partial} Y \Vert_{\rho - 2h} 
\leq && c h^{-\tau-d-1} \left(1+\vert \langle m 
\rangle \vert^{-1} \Vert m \Vert_\rho \right) \nonumber \\
&& \times \left( \Vert g\Vert_\rho +
c h^{-\tau-d-1} \Vert f\Vert_\rho \, 
\Vert m\Vert_\rho \right), \label{eqn:estDY2}
\end{eqnarray}
\begin{equation}
\vert a\vert \Omega^2 \leq \vert \langle m\rangle \vert^{-1} \left(
\Vert g\Vert_\rho+c h^{-\tau-d-1} 
\Vert f\Vert_\rho \, \Vert m\Vert_\rho \right).
\end{equation}
\end{lemme}
The proof is the same as the one for the estimates on $Z$.

   \subsection{Estimate on the new Hamiltonian}
      
The expression of the scalar functions of the Hamiltonian
expressed in the new actions and old angles are explicitly known, and
estimates on these functions are easy to obtain from the estimates
on the generating function. 
One has then to invert Eq.\ (\ref{eqn:tcan2}) in order to obtain 
the estimate on the Hamitonian expressed in the new angle variables.
The Jacobian of this transformation is 
\begin{equation}
\label{eqn:jac}
\mathrm{det} \left| \d\frac{\partial \phi_j'}{\partial \phi_k}
\right| = 1+\u{\Omega}\cdot\u{\partial}Y.
\end{equation}
In this section, we denote $V_\rho=\max (\Vert f\Vert_\rho, 
\Vert g\Vert_\rho)$ and $\Gamma=\max (1,\Vert m\Vert_\rho,
\vert \langle m \rangle \vert^{-1})$.\\
From Eq.\ (\ref{eqn:tcan2}) and estimate (\ref{eqn:estY}), one has
the following inequality:
\begin{equation}
\label{eqn:dphi}
\vert {\u \varphi}'-{\u \varphi} \vert \leq 2\Gamma^3c 
 h^{-\tau-d} \left(1+c h^{-\tau-d-1} \right) V_\rho,
\end{equation}
for all ${\u \varphi}\in \mathcal{D}_{\rho-2h}$
(we recall that $\Gamma\geq 1$). If we assume that $V_\rho$ and
$h$ are sufficiently small, one has an estimate on the new angles. More
precisely, we assume the following inequalities:
\begin{eqnarray}
  && \label{eqn:hyp1}
  \Gamma^3c ^2 h^{-2(\tau+d+1)}V_\rho \leq \frac{1}{4},\\
  && \label{eqn:hyp2}
  c h^{-\tau-d-1}\geq 1.
\end{eqnarray}
Then, for all ${\u \varphi}\in\mathcal{D}_{\rho-2h}$, we have 
\begin{equation}
\vert {\u \varphi}'-{\u \varphi}({\u \varphi}')\vert \leq h.
\end{equation}
As a consequence,
$\mathcal{D}_{\rho-2h}\subset \Phi(\mathcal{D}_{\rho-3h})$
by the map ${\u \varphi}'\mapsto {\u \varphi}=\Phi({\u \varphi}')$ given by
Eq.\ (\ref{eqn:tcan2}). In order to
express the estimates with respect to ${\u \varphi}'$, 
it suffices thus to restrict
the width of the strip $\mathcal{D}$ from $\rho-2h$ to $\rho-3h$, e.\ g.
$\Vert f'\Vert_{\rho-3h}\leq \Vert \tilde{f} \Vert_{\rho-2h}.$
Then, the estimates on the new perturbation $(g',f')$ are obtained from Eqs.\ 
(\ref{eqn:g})-(\ref{eqn:f}) and from the estimates on the generating function:
\begin{eqnarray}
&& 
\Vert g' \Vert_{\rho-3h} \leq 2^4 \Gamma^6 c^3 h^{-3(\tau +d+1)}
  V_\rho^2,\\
&& \Vert f' \Vert_{\rho-3h} \leq 2^3 \Gamma^5 c^2 h^{-2(\tau +d+1)}
  V_\rho^2.
\end{eqnarray}
Taking into account condition (\ref{eqn:hyp2}), we obtain the estimate 
on $V'_{\rho-3h}=\max(\Vert g' \Vert_{\rho-3h},\Vert f' \Vert_{\rho-3h})$:
\begin{equation}
\label{eqn:V'}
V'_{\rho-3h} \leq 2^4 \Gamma^6 c^3 h^{-3(\tau +d+1)} V_\rho^2.
\end{equation}
Concerning the quadratic term, we deduce
from Eqs.\ (\ref{eqn:estDY2}) and (\ref{eqn:m}) that
\begin{equation}
\Vert m'\Vert_{\rho-3h}\leq \Vert m\Vert_\rho \left(1+4\Gamma^3 c^2
h^{-2(\tau+d+1)}V_\rho \right)^2.
\end{equation}
Then using Hypothesis $(iii)$ of the theorem, we obtain 
$\Vert m'\Vert_{\rho-3h}\leq 2\Gamma$.\\
The mean value of $m'$ is determined by the integral
\begin{equation}
\langle m'\rangle=\int_{{\Bbb T}^d} \frac{d^d{\u \varphi}'}{(2\pi)^d} 
m'({\u \varphi}').
\end{equation}
With the change of variable ${\u \varphi}'\mapsto {\u \varphi}$ given by Eqs.\
(\ref{eqn:tcan2})-(\ref{eqn:jac}) and using Eq.\ (\ref{eqn:m}),
we rewrite the integral
\begin{eqnarray}
\langle m'\rangle&=&\int_{{\Bbb T}^d} \frac{d^d{\u \varphi}}{(2\pi)^d} 
\vert 1+{\u \Omega}\cdot{\u \partial} Y \vert \tilde{m}({\u \varphi}) 
\nonumber\\
&=&\int_{ {\Bbb T}^d} \frac{d^d{\u \varphi}}{(2\pi)^d} 
\vert 1+{\u \Omega}\cdot{\u \partial} Y \vert^3 m({\u \varphi}).\label{eqn:intm}
\end{eqnarray}
To estimate 
$\vert\langle m'\rangle \vert^{-1}$, we first estimate the difference
$\vert \langle m'\rangle-\langle m\rangle\vert$ using Eq.\ (\ref{eqn:intm}), and
we obtain
\begin{eqnarray}
\vert \langle m'\rangle-\langle m\rangle\vert \leq \Vert m \Vert_\rho &&
\left(3\Vert {\u \Omega}\cdot{\u \partial} Y \Vert_{\rho-2h}
+3\Vert {\u \Omega}\cdot{\u \partial} Y \Vert_{\rho-2h}^2\right. \nonumber \\
&& +\left. \Vert {\u \Omega}\cdot{\u \partial} Y \Vert_{\rho-2h}^3\right).
\end{eqnarray}
From Eq.\ (\ref{eqn:estDY2}) and condition (\ref{eqn:hyp1}), we have the 
estimate
\begin{equation}
\Vert {\u \Omega}\cdot{\u \partial} Y \Vert_{\rho-2h} \leq 4\Gamma^3 c^2
h^{-2(\tau+d+1)}V_\rho \leq 1,
\end{equation}
which leads to
\begin{equation}
\label{eqn:mm}
\vert \langle m'\rangle-\langle m\rangle\vert \leq 28\Gamma^4 
c^2 h^{-2(\tau+d+1)} V_\rho.
\end{equation}
One can easily check that from Hypothesis $(iii)$, 
 it follows that
\begin{equation}
\label{eqn:diffm}
\vert \langle m'\rangle-\langle m\rangle\vert \leq \frac{1}{2\Gamma}.
\end{equation}
Writing that $\langle m'\rangle=\langle m\rangle +\langle m'\rangle -
\langle m\rangle$, we have the following estimate on 
$\vert \langle m'\rangle\vert^{-1}$ :
\begin{equation}
\label{eqn:m-1}
\vert \langle m'\rangle\vert^{-1} \leq \frac{\vert \langle m\rangle\vert^{-1}}{
\left| 1-\vert \langle m\rangle\vert^{-1}
 \vert \langle m'\rangle-\langle m\rangle\vert \right|}.
\end{equation}
From Eq.\ (\ref{eqn:diffm}), we deduce that 
$\vert \langle m'\rangle\vert^{-1}\leq 2\Gamma$. Therefore, we have proved 
that $\Gamma'\equiv \max (1,\Vert m'\Vert_{\rho-3h},\vert \langle m'\rangle
\vert^{-1})\leq 2\Gamma$.


\section{Convergence of the iteration: KAM theorem}
\label{sec:conv}
   
The estimate (\ref{eqn:V'}) gives a precise meaning to the 
statement that the perturbation is reduced from
$V$ to $V^2$: we see that the actual reduction is somewhat smaller,
since $h$ depends on $V_\rho$ through the condition (\ref{eqn:hyp1}) [the
actual rate of reduction that  takes  this into account will be expressed by
Eq.\ (\ref{eqn:estvn}) below].
The counterpart is that the width of the 
domain of analyticity of
the new functions is reduced from $\rho$ to $\rho-3h$; for this
reason, $h$ is called the ``analyticity loss parameter''.
At the $n$th step of the iteration, we choose $h_n$ as the analyticity loss 
parameter such that the final domain of analyticity (after an infinite
number of iterations) does not shrink to zero. For instance, we choose
\begin{equation}
\label{eqn:ch1}
h_n=h 3^{-n+1}, \quad \mbox{ with  } h<\inf(2\rho/9,c^{1/(\tau+d+1)}).
\end{equation}
The second term in the inf is to guarantee also the condition (\ref{eqn:hyp2}).
We denote $\Gamma_n=\max (1, 
\Vert m^{(n)}\Vert_{\rho_n}, \vert\langle m^{(n)}\rangle \vert^{-1})$ and
$V_n= \max (\Vert g^{(n)}\Vert_{\rho_n},\Vert f^{(n)}\Vert_{\rho_n})$,
where $m^{(n)}$, $g^{(n)}$ and $f^{(n)}$ denote the three scalar functions
defining the Hamiltonian (\ref{hamiltonian}) after $n$ iterations,
 and 
\begin{equation}
\label{eqn:ch2}
\rho_n=\rho_{n-1}-3h_n=\rho-3\sum_{k=1}^n h_k.
\end{equation}
 The previous section gave the following estimates
\begin{equation}
 \label{eqn:Vn}
    V_n \leq 2^4 \Gamma_{n-1}^6 c^3 h_{n}^{-3(\tau+d+1)} V_{n-1}^2,
\end{equation}
\begin{eqnarray}
    \Vert m^{(n)}\Vert_{\rho_n}
      \leq && \Vert m^{(n-1)}\Vert_{\rho_{n-1}}\nonumber \\
      && \times \left( 1+4\Gamma_{n-1}^3c^2h_n^{-2(\tau+d+1)}
    V_{n-1}\right)^2.\label{eqn:mn}
\end{eqnarray}
We now prove that ${\Gamma_n}$ is bounded, and that ${V_n}$ tends
to zero faster than geometrically as $n$ goes to infinity.\\
Denoting
\begin{eqnarray}
&& V_0=\max(\Vert g \Vert_\rho,\Vert f \Vert_\rho),\\ 
&&  \Gamma=\max(1,\Vert m\Vert_\rho,|\langle m\rangle |^{-1}),\\ 
&& \gamma=2^{10} \Gamma^6 c^3 h^{-3(\tau+d+1)},\\
&& \delta=3^{3(\tau+d+1)},
\end{eqnarray}
we define the sequence of values 
\begin{equation}
\label{eqn:eps1}
\varepsilon_{n}=\gamma \delta^{n-1} \varepsilon_{n-1}^2
=\frac{(\gamma \delta V_0)^{2^n}}{\gamma \delta^{n+1}}.
\end{equation} 
We will show by induction that 
\begin{eqnarray}
&& \label{eqn:estgamn}
\Gamma_k \leq 2\Gamma,\\
&& \label{eqn:estvn}
V_k \leq \varepsilon_k,
\end{eqnarray}
for all $k$. If we assume $V_0$
to be sufficiently small, such that $\gamma \delta V_0  < 1$, 
i.\ e.
\begin{equation}
2^{10}\Gamma^6 c^3 h^{-3(\tau +d+1)}3^{3(\tau+d+1)}V_0 < 1,
\end{equation}
which is the hypothesis $(iii)$ of the theorem,
then $V_n$ tends to zero faster than  $\delta^{-n}=3^{-3n(\tau+d+1)}$ 
as $n$ goes to infinity.
The bounds (\ref{eqn:estgamn})-(\ref{eqn:estvn})
are satisfied for $k=0$, since $\Gamma_0=\Gamma>0$ and 
$V_0=\varepsilon_0$.\\
Suppose that for all $k<n$, these bounds are satisfied.
From Eqs.\ (\ref{eqn:Vn}) and (\ref{eqn:estgamn}), we deduce that
\begin{equation}
\label{eqn:vnest}
V_n\leq \gamma \delta^{n-1} V_{n-1}^2.
\end{equation} 
Then using Eq.\ (\ref{eqn:estvn}), we have $V_n \leq \varepsilon_n$.
Concerning $\Gamma_n$, Eq.\ (\ref{eqn:mn}) leads to
\begin{eqnarray*}
\Vert m^{(n)}\Vert_{\rho_n}\leq && \Vert m\Vert_{\rho} \\
&& \times \prod_{k=0}^{n-1}
\left(1+2^5\Gamma^3 c^2h^{-2(\tau+d+1)}3^{2k(\tau+d+1)}V_k\right)^2.
\end{eqnarray*}
Using Eq.\ (\ref{eqn:estvn}) and Hypothesis $(iii)$, we obtain
\begin{equation}
2^5\Gamma^3 c^2h^{-2(\tau+d+1)}3^{2k(\tau+d+1)}V_k\leq 3^{-(k+3)(\tau+d+1)}.
\end{equation}
Then we have
\begin{equation}
\Vert m^{(n)}\Vert_{\rho_n}\leq \Gamma \prod_{k=0}^{\infty}
\left(1+3^{-(k+3)(\tau+d+1)}\right)^2.
\end{equation}
Using the fact that $\prod (1+x_k)^2\leq \exp 2\sum x_k$, we show that the 
infinite product converges, and that it is smaller than $2$. Thus, 
$\Vert m^{(n)}\Vert_{\rho_n}\leq  2\Gamma$.\\
For $\vert \langle m^{(n)}\rangle \vert^{-1}$, we use estimate 
(\ref{eqn:m-1}):
\begin{equation}
\frac{\vert \langle m^{(n)}\rangle \vert^{-1}}{
\vert \langle m^{(n-1)}\rangle \vert^{-1}}\leq \frac{1}{\left| 1-
\vert \langle m^{(n-1)}\rangle \vert^{-1} 
\vert \langle m^{(n)}\rangle -  \langle m^{(n-1)}\rangle \vert \right|}.
\end{equation}
The difference $\vert \langle m^{(n)}\rangle -  \langle m^{(n-1)}\rangle \vert$
is evaluated as in the first step~(\ref{eqn:mm}):
\begin{eqnarray}
\vert \langle m^{(n)}\rangle -  \langle m^{(n-1)}\rangle \vert\leq &&
28\cdot 2^4\Gamma^4 c^2 h^{-2(\tau+d+1)} \nonumber \\
&& \times 3^{2(n-1)(\tau+d+1)} V_{n-1}.
\end{eqnarray}
Using Eq.\ (\ref{eqn:estvn}), Hypothesis $(iii)$, and the fact that
$\Gamma$ and $c h^{-(\tau+d+1)}$ are greater than one,
\begin{equation}
\vert \langle m^{(n)}\rangle -  \langle m^{(n-1)}\rangle \vert\leq 
2^{-1}\Gamma^{-1} 3^{-(n+2)(\tau+d+1)}.
\end{equation}
Then,
\begin{equation}
\vert \langle m^{(n)}\rangle \vert^{-1}\leq \Gamma \prod_{k=0}^\infty
\left(1-3^{-(k+2)(\tau+d+1)}\right)^{-1}.
\end{equation}
Using the fact that $\prod (1-x_k)^{-1} =\exp(-\sum \ln (1-x_k))$, and that
$-\ln(1-x)\leq x 2\ln 2$, $\forall x\in [0,1/2]$, we show by straightforward
calculations that $\vert \langle m^{(n)}\rangle \vert^{-1}\leq 2\Gamma$.\\
Thus we have proved (\ref{eqn:estgamn})-(\ref{eqn:estvn}) for all $k$.\\
We finally have to verify that with the choices (\ref{eqn:ch1})-(\ref{eqn:ch2}),
the conditions (\ref{eqn:hyp1})-(\ref{eqn:hyp2}) are satisfied at each step of
the iteration. Equation (\ref{eqn:hyp2}) is guaranteed by Eq.\ (\ref{eqn:ch1}):
$ch_n^{-(\tau+d+1)}\geq ch^{-(\tau+d+1)}\geq 1$.
For Eq.\ (\ref{eqn:hyp1}), we first notice that, since $\Gamma \geq 1$,
$\Gamma_n\leq 2\Gamma$, and
$c h^{-(\tau+d+1)} > 1$, we can write
$$
\Gamma_{n-1}^3 c^2 h_n^{-2(\tau+d+1)} < 2^{-7} \gamma \delta^n.
$$
Therefore, using Eq.\ (\ref{eqn:estvn}) and Hypothesis $(iii)$, we have
$$
\Gamma_{n-1}^3 c^2 h_n^{-2(\tau+d+1)} V_{n-1} < 2^{-7} 
\left(\gamma \delta V_0\right)^{2^{n-1}} <\frac{1}{4}.
$$
We have shown that, for sufficiently small initial perturbation, the
Hamiltonian converges under successive iterations of the transformation to
a Hamiltonian of the form (\ref{hamiltonian}) with $g=f=0$.
In order to complete the proof, we have still to verify that the composition
of the infinitely many canonical transformations is a well-defined finite
canonical transformation. It suffices to verify that 
$\vert {\u \varphi}^{(\infty)}-{\u \varphi} \vert$ and 
$\vert {\u A}^{(\infty)}-{\u A} \vert$ are finite. This is an immediate 
consequence of the fast convergence of the iteration. 
From Eq.\ (\ref{eqn:dphi}), we deduce e.\ g.\ that
\begin{equation}
\vert {\u \varphi}^{(\infty)}-{\u \varphi} \vert \leq \sum_{n=1}^\infty
h_n =3h/2.
\end{equation}
Analogously, using Eqs.\ (\ref{eqn:tcan1}), (\ref{eqn:estDZ2}),
(\ref{eqn:estDY1}), (\ref{eqn:estDY2}), and (\ref{eqn:estvn}), for
$\vert {\u A}\vert \leq A_0$, we can bound
\begin{equation}
\vert {\u A}^{(\infty)}-{\u A} \vert <\mbox{const }\times \sum_{n=1}^\infty
h_n.
\end{equation}

\section*{acknowledgments}
We thank G. Benfatto, L. H. Eliasson, G. Gallavotti, 
M. Herman, and H. Koch for helpful discussions.
Support from EC contract No.\ ERBCHRXCT94-0460 for the project
``Stability and universality in classical mechanics''
and from  the Conseil R\'egional de Bourgogne is acknowledged.



\newpage

\begin{references}

\bibitem{kolmogorov}
A.~N. Kolmogorov, Dokl. Akad. Nauk SSSR {\bf 98}, 527 (1954)
[in {\em Stochastic behaviour in classical and quantum
Hamiltonian systems}, edited by G. Casati and J. Ford,
Lecture Notes in Physics Vol. 93 (Springer,Berlin,1979), p. 51].

\bibitem{arnold}
V.~I. Arnold, Russ. Math. Surv. {\bf 18}, 9 (1963).

\bibitem{moser}
J.~Moser, {Nachr. Akad. Wiss. Goett.}, Math.-Phys. Kl. IIa {\bf 1},1 (1962).

\bibitem{gallavotti3}
G.~Gallavotti, in {\em Scaling and {S}elf-{S}imilarity in {P}hysics},
edited by J. Fr\"ohlich (Birkh\"auser, Boston, 1983).

\bibitem{govin}
M. Govin, C. Chandre, and H. R. Jauslin, Phys. Rev. Lett. {\bf 79}, 3881 (1997).

\bibitem{chandre}
C. Chandre, M. Govin, and H. R. Jauslin, Phys. Rev. E {\bf 57}, 1536 (1998).

\bibitem{cgjk}
C. Chandre, M. Govin, H. R. Jauslin, and H. Koch, Phys. Rev. E
{\bf 57}, 6612 (1998).

\bibitem{thirring}
W.~Thirring, {\em A Course in Mathematical Physics I: Classical Dynamical
  Systems} (Springer-Verlag, Berlin, 1992), p. 153.
  
\bibitem{jauslin}
H. R. Jauslin, in {\em II Granada Lectures in Computational Physics},
edited by P. L. Garrido and J. Marro (World Scientific, Singapore, 1993).
  
\bibitem{arnold2}
V.~I. Arnold, Russ. Math. Surv. {\bf 18}, 85 (1963).

\bibitem{chengsun}
Ch.-Q.~Cheng and Y.-S. Sun, J. Differential Equations {\bf 114}, 288 (1994).

\bibitem{russmann}
H. R\"ussmann, in \textit{Stochastics, Algebra and Analysis in Classical and
Quantum Dynamics}, edited by S. Albeverio, Ph. Blanchard and D. Testard, 
Mathematics and its Applications, Vol. 59 (Kluwer, Dordrecht, 1990).

\bibitem{gallavotti1}
G.~Gallavotti, Commun. Math. Phys. {\bf 164}, 145 (1994).

\bibitem{gallavotti2}
G.~Gallavotti, Rev. Math. Phys. {\bf 6}, 343 (1994).
  
\bibitem{paul}
R.~D. Paul, archived in mp\_arc@math.utexas.edu, \#98-9 (1998).

\bibitem{delshams}
A. Delshams and P. Guti\'errez, J. Differential Equations {\bf 128}, 415 (1996).

\bibitem{broer}
H. W. Broer and G. B. Huitema, J. Differential Equations {\bf 90}, 52 (1991).

\bibitem{goldstein}
H. Goldstein, {\em Classical Mechanics} (Addison-Wesley, Reading, Mass., 1980).

\bibitem{gallavottiL}
G. Gallavotti, {\em The Elements of Mechanics} (Springer-Verlag, New York,
1983).

\bibitem{benettin}
G. Benettin, L. Galgani, A. Giorgilli, and J. M. Strelcyn, 
Nuovo~Cimento~{\bf 79}B,~201~(1984).

\end{references}
\end{document}